# ON THE PARAMETER ESTIMATION OF ARMA(p,q) MODEL BY APPROXIMATE BAYESIAN COMPUTATION

LINGHUI LI[1], ANSHUI LI[2], and HUIZENG ZHANG[3*]

[1]Department of mathematics, school of science, Hangzhou Normal University, Hangzhou Zhejiang, P.R. China

[2]Department of mathematics, school of science, Hangzhou Normal University, Hangzhou Zhejiang, P.R. China

[3]Department of mathematics, school of science, Hangzhou Normal University, Hangzhou Zhejiang, P.R. China

*Corresponding author

**Abstract.** In this paper, the parameter estimation of *ARMA(p,q)* model is given by approximate Bayesian computation algorithm. In order to improve the sampling efficiency of the algorithm, approximate Bayesian computation should select as many statistics as possible with parameter information in low dimension. Firstly, we use the autocorrelation coefficient of the first *p+q* order sample as the statistic and obtain an approximate Bayesian estimation of the *AR* coefficient, transforming the *ARMA(p,q)* model into the *MA(q)* model. Considering the first *q* order sample autocorrelation functions and sample variance as the statistics, the approximate Bayesian estimation of *MA* coefficient and white noise variances can be given. The method mentioned above is more accurate and powerful than the maximum likelihood estimation, which is verified by the numerical simulations and experiment study.

## 1. INTRODUCTION

Maximum likelihood estimation is one common parameter estimation method for time series *ARMA(p,q)* model. This method is relatively complex with large estimation error in high-order case. Graupe et.al proposed to fit the *ARMA* model into a higher order *AR* model by the least square method [1]. By solving the compatible algebraic equations, the parameter estimates of the *ARMA* model can be obtained, but the accuracy of the estimations is not high. In order to improve the accuracy, two-stage least squares method is given by Z Deng et al [2], but the convergence rate was relatively slow. Based on affine geometry algorithm [3], L.Shen obtained accurate linear and non-linear *ARMA* model parameter estimation.

Approximate Bayesian computation (*ABC* in short) is a novel algorithm in Bayesian statistic, sampling the approximate posterior distribution of parameters with rejection method. Rubin first gave the basic idea of approximate Bayesian [6] in 1984, Tavaré et. al brought the algorithm into statistical inference of DNA data [7] later. Pritchard and others extended this algorithm [8] and used the extended version of *ABC* algorithm to study the variation of human *Y* chromosome. Approximate Bayesian computation, regarded as a powerful method of parameter estimation for complex models, has been widely used in population genetics, systems biology, epidemiology and systematic geography [9,10] since 2003. In order to improve the sampling efficiency of the *ABC* algorithm, sufficient statistics of parameters should be selected. It is generally difficult to obtain sufficient statistics of parameters in most cases. Jean-Michel Marin et.al proposed the *ABC* method to estimate the parameters of the *MA(q)* model [11].

In this paper, the autocovariance function is selected as the statistic, and the estimation accuracy is not high when the white noise variance is known. According to the relationship between the autocorrelation function and the model coefficient of *ARMA(p,q)* model, this paper first obtain the estimation of the *AR* coefficient with *ABC* algorithm, transforming the *ARMA(p,q)* model into the *MA(q)* model. We then take the *q* order sample autocorrelation function and the sample variance as statistics, and get the estimation of the coefficient of *MA(q)* model and the variance of the white noise. Our method greatly improves the accuracy of estimation for the *ARMA(p,q)* model. In other words, this paper implement a simple and powerful approximate Bayesian algorithm to estimate *ARMA(p,q)* model parameters.

This paper is organized as follows: some necessary techniques of *ARMA(p,q)* model and the *ABC* algorithm which will be used in the sequel are briefly described in Section 2 and Section 3 respectively; our main results are shown in Section 4; simulations and experimental study will be given in Section 5 and Section 6.

## 2. ARMA(p,q) MODEL

Let us begin with some basic definitions and properties of the *ARMA(p,q)* model which will used in the sequel. Most of the definitions and properties list below can be found in [12].

**Definition 1.** Let $\{Y_t : t \in T\}$ be a time series in which $T=\{0, \pm 1, \pm 2, \cdots\}$, if $\forall t \in T$,

$$Y_t = \mu + e_t - \theta_1 e_{t-1} - \theta_2 e_{t-2} - \cdots - \theta_q e_{t-q},$$

we call such a series a moving average model of order *q* and abbreviate the name to *MA(q)*, where $\theta_q \neq 0$, $e_t$ is a white noise series. The series $\{Y_t : t \in T\}$ is called the centralized *MA(q)* model when $\mu = 0$.

Let $\Theta(x) = 1 - \theta_1 x - \theta_2 x^2 - \cdots - \theta_q x^q$, the $\Theta(x) = 0$ is called the characteristic equation of the *MA(q)* model, the *MA(q)* model is reversible if and only if the root of the characteristic equation $\Theta(x) = 0$ is outside the unit circle.

**Proposition 1.** If *MA(q)* process is reversible, then $\theta_i$ satisfies: $|\theta_i| < C_q^i, i = 1, 2, \cdots, q$.

**Proof:** Assume characteristic equation is

$$\Theta(x) = 1 - \theta_1 x - \theta_2 x^2 - \cdots - \theta_q x^q = (1 - \lambda_1 x)(1 - \lambda_2 x)(1 - \lambda_3 x) \cdots (1 - \lambda_q x)$$

the right-hand term of the upper formula is expanded

$$1 - \left(\sum_{i=1}^q \lambda_i\right) x + \left(\sum_{i<j} \lambda_i \lambda_j\right) x^2 - \left(\sum_{i<j<k} \lambda_i \lambda_j \lambda_k\right) x^3 + \cdots + (-1)^q \left(\prod_{i=1}^q \lambda_i\right) x^q,$$

compared with the left-hand formula

$$\theta_1 = \sum_{i=1}^q \lambda_i, \theta_2 = -\sum_{i<j} \lambda_i \lambda_j, \theta_3 = \sum_{i<j<k} \lambda_i \lambda_j \lambda_k, \cdots, \theta_q = (-1)^q \left(\prod_{i=1}^q \lambda_i\right).$$

for the *MA(q)* process is reversible, the root of the equation $\theta_i = \dfrac{1}{\lambda_i}$ is outside the unit circle, $i=1,2,\cdots,q$, so $|\lambda_i|<1, i=1,2,\cdots q$. Then $|\theta_1|<q=C_q^1, |\theta_2|<C_q^2, |\theta_3|<C_q^3, \cdots, |\theta_q|<C_q^q$, thus $|\theta_i|<C_q^i (1\le i \le q)$.

**Proposition 2.** The relationship between the autocorrelation coefficient $\rho_k$ of *MA(q)* model and $\theta_i(1\le i \le q)$ is as follows:

$$\rho_k = \begin{cases} 1, k=0 \\ \dfrac{-\theta_k + \sum\limits_{i=1}^{q-k}\theta_i \theta_{k+i}}{1+\theta_1^2 + \cdots + \theta_q^2}, 1\le k \le q \\ 0, k>q \end{cases}$$

**Definition 2.** Let $\{Y_t : t\in T\}$ be a time series in which T=$\{0, \pm 1, \pm 2, \cdots\}$, if $\forall t\in T$,

$$Y_t = c + \phi_1 Y_{t-1} + \phi_2 Y_{t-2} + \cdots + \phi_p Y_{t-p} + e_t$$

we call such a series an autoregressive model of order *p* and abbreviate the name to *AR(p)*, where, $\phi_p \neq 0$, $\{e_t\}$ is a white noise series. The series $\{Y_t : t\in T\}$ is called the centralized *AR(p)* model when c=0.

Let $\Phi(x) = x^p - \phi_1 x^{p-1} - \phi_2 x^{p-2} - \cdots - \phi_p$, the $\Phi(x)=0$ is called the characteristic equation of the *AR(p)* model, the *AR(p)* model is stationary if and only if the root of the characteristic equation $\Phi(x)=0$ is in the unit circle.

**Proposition 3.** If *AR(p)* process is stationary, then $\phi_i$ satisfies: $|\phi_i|<C_p^i, i=1,2,\cdots,p$.

**Proof:** Similar to the proof of *Proposition 1*, we will not repeat it here.

**Definition 3.** Let $\{Y_t : t\in T\}$ be a time series in which T=$\{0, \pm 1, \pm 2, \cdots\}$, if $\forall t\in T$,

$$Y_t = c + \phi_1 Y_{t-1} + \phi_2 Y_{t-2} + \cdots + \phi_p Y_{t-p} + e_t - \theta_1 e_{t-1} - \theta_2 e_{t-2} - \cdots - \theta_q e_{t-q}$$

We call such a series an autoregressive moving average model and abbreviate the name to *ARMA(p,q)*, where $\phi_p \neq 0, \theta_q \neq 0, \{e_t\}$ is a white noise series and $e_t \sim N(0,\sigma^2), \sigma^2$ is the variance of the white noise. The series $\{Y_t : t\in T\}$ is called the centralized *ARMA(p,q)* model when c=0. The *ARMA(p,q)* model studied in this paper refers to the centralized *ARMA(p,q)* model.

Obviously, the *ARMA(p,q)* model degenerates into the *MA(q)* model when p=0; it analogously degenerates into a *AR(p)* model when q=0. So the *AR(p)* and *MA(q)* model are special cases of the

ARMA(p,q) model, and the statistical properties of the ARMA(p,q) model are also a combination of the AR(p) and MA(q) model.

**Proposition 4.** If ARMA(p,q) model is stationary, then $\phi_i$ satisfies: $|\phi_i| < C_p^i, i = 1, 2, \cdots, p$.

**Proof:** If ARMA(p,q) model is stationary, then the root of the equation $\Phi(x) = 0$ is in the unit circle. According to the *Proposition 3*, $|\phi_i| < C_p^i, i = 1, 2, \cdots, p$.

**Proposition 5.** If ARMA(p,q) model is reversible, then $\theta_i$ satisfies: $|\theta_i| < C_q^i, i = 1, 2, \cdots, q$.

**Proof:** If ARMA(p,q) model is reversible, then the root of equation $\Theta(x) = 0$ is outside the unit circle. According to the *Proposition 1*, $|\theta_i| < C_q^i, i = 1, 2, \cdots, q$.

**Proposition 6.** The relationship between the autocorrelation coefficient $\rho_k$ of stationary ARMA(p,q) model and $\phi_i (1 \le i \le p)$ is as follows:

$$\begin{cases} \rho_{p+1} = \phi_1 \rho_p + \phi_2 \rho_{p-1} + \cdots + \phi_p \rho_1 \\ \rho_{p+2} = \phi_1 \rho_{p+1} + \phi_2 \rho_p + \cdots + \phi_p \rho_2 \\ \cdots \cdots \\ \rho_{p+q} = \phi_1 \rho_{p+q-1} + \phi_2 \rho_{p+q-2} + \cdots + \phi_p \rho_p \end{cases}$$

If two stationary ARMA(p,q) models have the same autocorrelation of order $p+q$, and

$$\begin{bmatrix} \rho_p & \rho_{p-1} & \cdots & \rho_1 \\ \rho_{p+1} & \rho_p & \cdots & \rho_2 \\ \cdots & \cdots & \cdots & \cdots \\ \rho_{p+q-1} & \rho_p & \cdots & \rho_q \end{bmatrix}$$

is reversible, the AR coefficients of the two ARMA(p,q) models are the same.

The above propositions are well-known properties, and no proof is given here.

## 3. APPROXIMATE BAYESISN COMPUTATION

In order to illustrate the approximate Bayesian computation method, let's start with the Bayesian model. Assume the density of $X = (X_1, X_2, \cdots, X_n)^T$ is $p_{X|\theta}(x|\theta)$, where $\theta$ is the parameter of model, $x = (x_1, x_2, \cdots, x_n)^T$. $\theta$ has a prior density of $p_\theta(\theta)$, in the case of $X = x$ the posterior density of $\theta$ is

$$p_{\theta|X}(\theta|x) = \frac{p_{X|\theta}(x|\theta) p_\theta(\theta)}{\int p_{X|\theta}(x|t) p_\theta(t) dt}$$

According to the above formula, the calculation of posterior density involves the calculation of likelihood function $p_{X|\theta}(x|\theta)$ and $\int p_{X|\theta}(x|t) p_\theta(t) dt$. In complex models, this is very difficult.

Approximate Bayesian computation can avoid these two problems. Here we give a brief introduction.

Approximate Bayesian computation, abbreviated as *ABC*, is a sampling simulation method of approximate posterior distribution. The most remarkable feature of this method is that it's easy to implement and does not need to calculate the likelihood function $p_{X|\theta}(x|\theta)$. Generally speaking, it is difficult to calculate the likelihood function and posterior density in many cases, while there is no specific or closed expression. In order to estimate the parameters of the complex statistical model, we can use the *ABC* algorithm to simulate posterior samples and replace the calculation of likelihood function with the information of these samples to obtain the estimation of model parameters.

First, we give the basic *ABC* algorithm.

**Algorithm 1.** (Basic *ABC* algorithm)

input: samples $x^* \in R^n$; the prior distribution $p_\theta(\cdot)$ of $\theta$; likelihood function $p_{X|\theta}(\cdot|\theta)$;

the distance $\rho(\cdot,\cdot)$ of $R^n$; an approximation parameter $\delta > 0$.

output: samples $(\theta_1, \theta_2, \cdots, \theta_n)$ of approximate posterior distribution $p_{\theta|X}(\theta|x^*)$.

algorithm steps:
*for i=1 to n do*
*repeat*

generation $\theta'$ based on prior distribution $p_\theta(\cdot)$;

generation $x$ based on likelihood function $p_{X|\theta}(\cdot|\theta')$.

*until* $\rho(x, x^*) \le \delta$

$\theta_i = \theta'$

*end for*

The disadvantage of the basic *ABC* algorithm is that the probability of rejection of $\theta$ is very high when the dimension of *n* is large, which greatly increases the computational difficulty. To solve this issue, the low-dimensional sufficient statistic $S(x) \in R^d$ of parameter $\theta$ instead of *x* (*d<<n*) is used, which leads to an improved *ABC* algorithm.

**Algorithm 2.** (Improved *ABC* algorithm)

input: samples $x^* \in R^n$; the prior distribution $p_\theta(\cdot)$ of $\theta$; likelihood function $p_{X|\theta}(\cdot|\theta)$;

the distance $d(\cdot,\cdot)$ of $R^d$; an approximation parameter $\delta > 0$.

output: samples $(\theta_1, \theta_2, \cdots, \theta_n)$ of approximate posterior distribution $p_{\theta|X}(\theta|x^*)$.

algorithm steps:
*for i=1 to n do*
*repeat*

generation $\theta'$ based on prior distribution $p_\theta(\cdot)$;

generation $x$ based on likelihood function $p_{X|\theta}(\cdot|\theta')$.

*until* $d(S(x), S(x^*)) \leq \delta$

$\theta_i = \theta'$

*end for*

It can be proved that the samples obtained is from posterior distribution when $\delta \to 0$.

## 4. *ABC* ALGORITHMS OF THE PARAMETER ESTIMATION OF *ARMA(p,q)* MODEL

In this section, we will give our main algorithms for the parameters estimations of *ARMA(p,q)* model.

Let $\{Y_t : t \in T\}$ is a stationary *ARMA(p,q)* model,

$$Y_t = c + \phi_1 Y_{t-1} + \phi_2 Y_{t-2} + \cdots + \phi_p Y_{t-p} + e_t - \theta_1 e_{t-1} - \theta_2 e_{t-2} - \cdots - \theta_q e_{t-q}$$

where $T = \{0, \pm 1, \pm 2, \cdots\}$, $\phi_p \neq 0$, $\theta_q \neq 0$, $\{e_t\}$ is white noise series, and $e_t \sim N(0, \sigma^2)$, $\sigma^2$ is the variance of the white noise.

In order to guarantee the uniqueness of the model coefficient estimates, we assume that the model *ARMA(p,q)* is reversible.

Let $\phi = (\phi_1, \phi_2, \cdots, \phi_p)$, $\theta = (\theta_1, \theta_2, \cdots, \theta_q)$,

$D_1 = \{\phi :$ the root of $x^p - \phi_1 x^{p-1} - \phi_2 x^{p-2} - \cdots - \phi_p = 0$ is in the unit circle$\}$,

$D_2 = \{\theta :$ the root of $1 - \theta_1 x - \theta_2 x^2 - \cdots - \theta_q x^q = 0$ is outside the unit circle$\}$.

Obviously, the *ARMA(p,q)* model is stationary and reversible if and only if $\phi \in D_1, \theta \in D_2$.

In this paper, we assume that the prior distribution of $\phi$ is a uniform distribution on $D_1$, the prior distribution of $\theta$ is a uniform distribution on $D_2$, the prior distribution of $\sigma$ is inverse Gamma distribution, that is $\sigma = \frac{1}{\tau}$, $\tau \sim \Gamma(\alpha, \beta)$, $\alpha, \beta$ is a parameter of gamma distribution and is known. $\tau, \theta, \phi$ is independence from one another.

In the *ABC* algorithm, the first step is to generate a uniformly distributed $\phi$ on $D_1$ and $\theta$ on $D_2$. According to the **Proposition 4**, if the *ARMA(p,q)* model is stationary, we have $\phi \in \Delta_1 = \{\phi \in \Re^p : |\phi_i| < C_p^i, i = 1, 2, \cdots, p\}$. Thus we first simulate a random vector $\phi$ which obeys the uniform distribution on $\Delta_1$, then determine whether $\phi$ satisfies {the root of $x^p - \phi_1 x^{p-1} - \phi_2 x^{p-2} - \cdots - \phi_p = 0$ is in the unit circle}, if satisfied then accept it, else refuse it. Thus an algorithm for simulating uniformly distributed random vectors on $D_1$ is obtained. According to the **Proposition 5**, if ARMA(p,q) model is reversible, then $\theta \in \Delta_2 = \{\theta \in \Re^q : |\theta_i| < C_p^i, i = 1, 2, \cdots, q\}$. Similarly, we first simulate a random vector $\theta$, which obeys the uniform distribution on $\Delta_2$, then determine whether $\theta$ is satisfied {the root of $1 - \theta_1 x - \theta_2 x^2 - \cdots - \theta_q x^q = 0$ is outside the unit circle}, if satisfied then accept it, else refuse it. Thus an algorithm for simulating uniformly distributed random vectors on $D_2$ is obtained as follows.

**Algorithm 3.**
input: order p,q of the model

output: generate a random vector $\phi$ that obeys the uniform distribution on $D_1$;

generate a random vector $\theta$ that obeys the uniform distribution on $D_2$.

Algorithm steps:
1. *repeat*

generate a random vector $\phi$ that obeys the uniform distribution on $\Delta_1$

*until* $\phi$ satisfies: {the root of $x^p - \phi_1 x^{p-1} - \phi_2 x^{p-2} - \cdots - \phi_p = 0$ is in the unit circle}

output $\phi$

2. *repeat*

generate a random vector $\theta$ that obeys the uniform distribution on $\Delta_2$

*until* $\theta$ satisfies: {the root of $1 - \theta_1 x - \theta_2 x^2 - \cdots - \theta_q x^q = 0$ is outside the unit circle}

output $\theta$

According to the **Proposition 6**, if two stationary ARMA(p,q) models have the same autocorrelation of order p+q, and

$$\begin{bmatrix} \rho_p & \rho_{p-1} & \cdots & \rho_1 \\ \rho_{p+1} & \rho_p & \cdots & \rho_2 \\ \cdots & \cdots & \cdots & \cdots \\ \rho_{p+q-1} & \rho_p & \cdots & \rho_q \end{bmatrix}$$

is reversible, the AR coefficients of the two ARMA(p,q) models are the same. So we use the autocorrelation coefficient of the first order p+q sample as statistics and use the ABC algorithm to estimate $\phi = (\phi_1, \phi_2, \cdots, \phi_p)$, the specific algorithm is as follows.

**Algorithm 4.**

input: samples $y^* = (y_1^*, y_2^*, \cdots, y_n^*)$; order $p,q$ of the model; sample size $N$; an approximation parameter $\varepsilon$; the distance $d(\cdot,\cdot)$; the variance of white noise series $\sigma^2$.

output: approximate Bayesian estimation $\hat{\phi}$ of $\phi$.

Algorithm steps:

1. compute the autocorrelation coefficient of sample $y^*$, $\hat{\rho}(y^*) = (\hat{\rho}_1^*(y), \hat{\rho}_2^*(y), \cdots, \hat{\rho}_{p+q}^*(y))$.
2. *for i=1 to N do*
*repeat*

    according to the *Algorithm 3*, random generation $(\phi, \theta)$

    random generation the time Series $y = (y_1, y_2, \cdots, y_n)$ of *ARMA(p,q)* model with coefficient of $(\phi, \theta)$, white noise variance of $\sigma^2$, calculating the autocorrelation coefficient of the series $\hat{\rho}(y) = (\hat{\rho}_i(y), i=1, 2, \cdots, p+q)$.

*until* $d(\hat{\rho}(y), \hat{\rho}(y^*)) \leq \varepsilon$

$\phi_i = \phi$

*end for*

3. approximate Bayesian estimation $\hat{\phi}$ of $\phi$, $\hat{\phi} = \frac{1}{N}\sum_{i=1}^{N}\phi_i$.

Let $X_t = Y_t - \phi_1 Y_{t-1} - \phi_2 Y_{t-2} - \cdots - \phi_p Y_{t-p}$, then

$$X_t = e_t - \theta_1 e_{t-1} - \theta_2 e_{t-2} - \cdots - \theta_q e_{t-q}$$

is a *MA(q)* model. Assume $\hat{\phi}$ is the estimate of $\phi$ in *Algorithm 4*, bring samples $y^* = (y_1^*, y_2^*, \cdots, y_n^*)$ into $X_t = Y_t - \hat{\phi}_1 Y_{t-1} - \hat{\phi}_2 Y_{t-2} - \cdots - \hat{\phi}_p Y_{t-p}$, t=p+1,p+2,$\cdots$,n. We get the samples $\{x_i^* : p+1 \leq i \leq n\}$ of $X_t$. According to the **Proposition 2**, The model coefficient $\theta$ of *MA(q)* process is determined by the autocorrelation coefficient of the first $q$ order, which is independent of the variance of white noise. So when estimating the parameter $(\theta, \sigma^2)$, take the autocorrelation coefficient of the sample's first $q$ order as statistics, and estimated by *ABC* algorithm $\theta = (\theta_1, \theta_2, \cdots, \theta_q)$, then an estimate of $\sigma^2$ is given by using sample variance as a statistic. The specific algorithm is as follows.

**Algorithm 5.**

input: samples $x^* = (x^*_{p+1}, x^*_{p+2}, \cdots, x^*_n)$, where $x^*_{p+i} = y^*_{p+i} - \phi_1 y^*_{p+i-1} - \phi_2 y^*_{p+i-2} - \cdots - \phi_p y^*_i$

$(i = 1, 2, \cdots, n-p)$; order $q$ of the model; parameters $\alpha, \beta$ of Gamma distribution; sample size $N$; an approximation parameter $\varepsilon$; the distance $d(\cdot, \cdot)$ of $R^d$.

output: approximate Bayesian estimation $\hat{\theta}, \hat{\sigma}^2$ of $\theta, \sigma^2$.

Algorithm steps:

1. compute the autocorrelation coefficient of sample $x^*$, $\hat{\rho}(x^*) = (\hat{\rho}_1^*(x), \hat{\rho}_2^*(x), \cdots, \hat{\rho}_q^*(x))$

2. *for i=1 to N do*

*repeat*

    according to the *Algorithm 3*, random generation $\theta$

    random generation the time series $x = (x_1, x_2, \cdots, x_{n-q})$ of *MA(q)* model with coefficient of $\theta$, white noise variance of $\sigma^2$, calculating the autocorrelation coefficient of the series

    $\hat{\rho}(x) = (\hat{\rho}_i(x), i = 1, 2, \cdots, q)$

*until* $d(\hat{\rho}(x), \hat{\rho}(x^*)) \leq \varepsilon_1$

$\theta_i = \theta'$

*end for*

3. compute approximate Bayesian estimation of $\theta$, $\hat{\theta} = \dfrac{1}{N} \sum\limits_{i=1}^{N} \theta_i$.

4. compute the variance $\sigma^{*2}$ of samples $x^*$.

5. *for i=1 to N do*

*repeat*

    generating random numbers $\tau \sim \Gamma(\alpha, \beta)$. Let $\sigma = \dfrac{1}{\tau}$,

    random generation the time series $x = (x_1, x_2, \cdots, x_{n-q})$ of *MA(q)* model with coefficient of $\hat{\theta}$, white noise variance of $\sigma^2$, calculating the variance $\sigma^2$ of samples;

*until* $\left| \sigma^{*2} - \sigma^2 \right| \leq \varepsilon_2$

$\sigma_i = \sigma$

*end for*

6. compute approximate Bayesian estimation for the variance of white noise $\hat{\sigma}^2 = \dfrac{1}{N} \sum\limits_{i=1}^{N} \sigma_i^2$.

## 5. NUMERICAL SIMULATION

To illustrate the effect of *ABC* parameter estimation, we choose *ARMA(2,2)* model for numerical simulation. Let a *ARMA(2,2)* model

$$Y_t = 0.6Y_{t-1} + 0.2Y_{t-2} + e_t - 0.3e_{t-1} - 0.4e_{t-2}, t \in \{0, \pm 1, \pm 2, \cdots\}$$

where white noise series $e_t \sim N(0, 2^2)$.

Firstly, time series samples $x^* = (x_1^*, x_2^*, \cdots x_{1000}^*)$ are generated according to the model above. According to the *Algorithm 4*, simulated generated samples $\phi$ with size of *100000*, and the threshold parameter is $\varepsilon = 0.0005$. We will choose the autocorrelation coefficient as parameter statistic, and the distance is ranked from small to large. The first *50* $\phi$ of the smallest distance is taken as the posterior sample, and the average value of the posterior sample is taken to obtain the estimated value of $\phi$. Then by *Algorithm 5*, we generate a sample of $\theta$ with size of *100000*, with threshold parameter $\varepsilon = 0.0003$. The autocorrelation coefficient is used as parameter statistic and we take the nearest posterior sample of *30* $\theta$, which is used to estimate the parameter $\theta$.

We also select a threshold parameter $\varepsilon = 0.0001$ and a posteriori sample of $\sigma$ with size of *10* was simulated with the mean of this sample as the estimate of $\sigma$. *Figure 1* and *Figure 2* respectively represent the sample scatter plots of $\phi = (\phi_1, \phi_2)$ and $\theta = (\theta_1, \theta_2)$ approximated by the *ABC* algorithm. *Figure 3* and *Figure 4* respectively represent histograms of $\phi$ and $\theta$ for posterior distribution sampling. To verify the effectiveness of the *ABC* algorithm, we compare our results with the maximum likelihood method. The results are shown in *Table 1*. In conclusion, the sample points obtained by the *ABC* algorithm are closely concentrated in the vicinity of $\phi = (0.6, 0.2)$, $\theta = (0.3, 0.4)$, which is more accurate than maximum likelihood estimation.

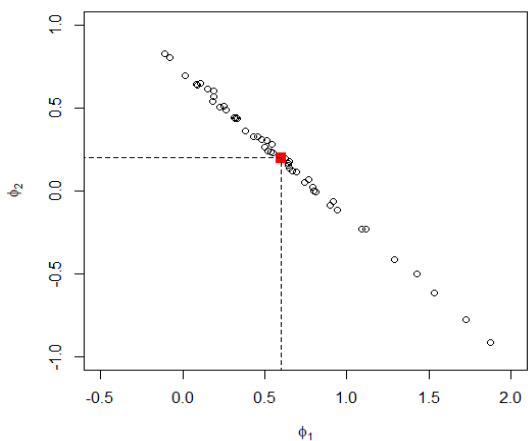

Figure 1

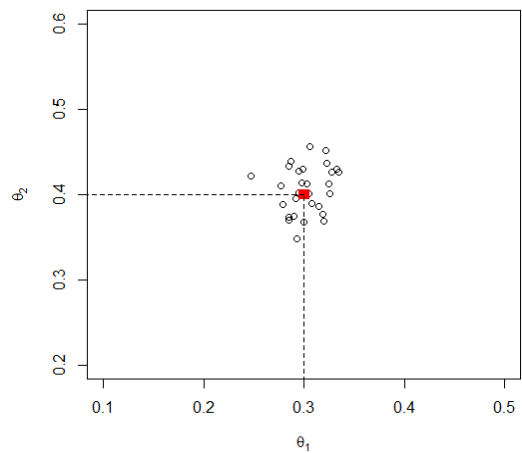

Figure 2

Figure 2. The sample scatter plots of $\phi$ approximated by the ABC algorithm

Figure 2. The sample scatter plots of $\theta$ approximated by the ABC algorithm

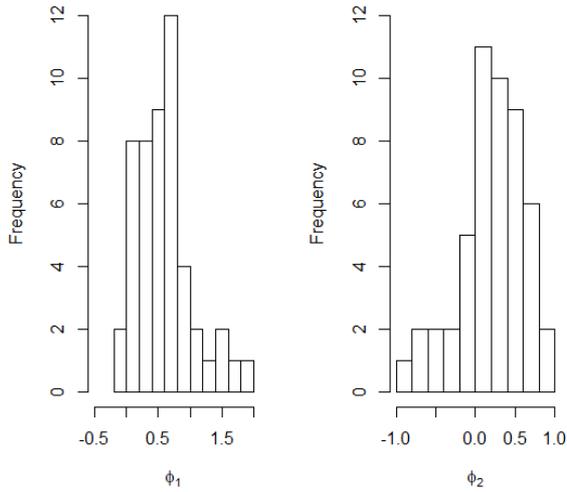 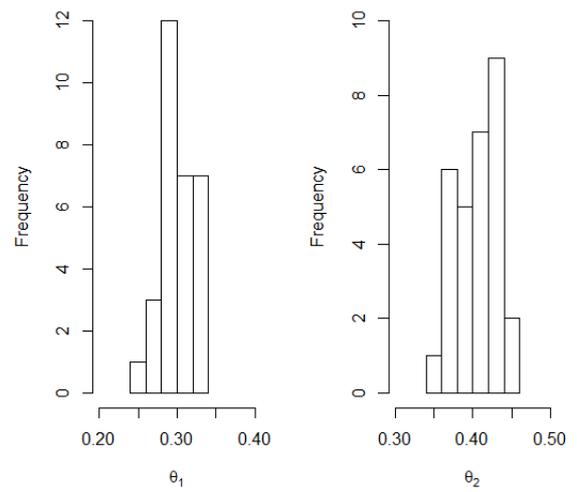

Figure 3                                    Figure 4

Figure 3. histograms of $\phi$ for posterior distribution sampling

Figure 4. histograms of $\theta$ for posterior distribution sampling

Table 1. Comparison of the maximum likelihood estimation and ABC algorithm

|  | Maximum Likelihood Estimation | | ABC Estimation | |
| --- | --- | --- | --- | --- |
|  | Estimation | Relative Error | Estimation | Relative Error |
| $\phi_1$ | 0.6236 | 3.93% | 0.5932088 | 1.13% |
| $\phi_2$ | 0.1783 | 10.85% | 0.2035063 | 1.75% |
| $\theta_1$ | 0.2729 | 9.03% | 0.3015688 | 0.52% |
| $\theta_2$ | 0.4034 | 0.85% | 0.4053529 | 1.34% |
| $\sigma^2$ | 3.949 | 1.28% | 3.953303 | 1.17% |

## 6. EXAMPLE ANALYSIS

The primary industry is an industry that produces natural objects and plays an important role in daily life. *Figure 5* gives the time series diagram of the year-by-year growth of the primary industry. With *ADF* test of this sequence, we can show that the sequence is approximate stationary. We can select the alternative model through the *ACF* chart and *PACF* chart, such as *Figure 6*, and rank the model as *ARMA(1,1)*. We take the year-by-year growth of the first industry from the first quarter in 2006 to the first quarter in 2018 as experiment data. Using the *ARMA(1,1)* model to forecast the quarters of 1~2,1~3,1~4 in 2018, and then comparing with the actual value to illustrate the effect of our model.

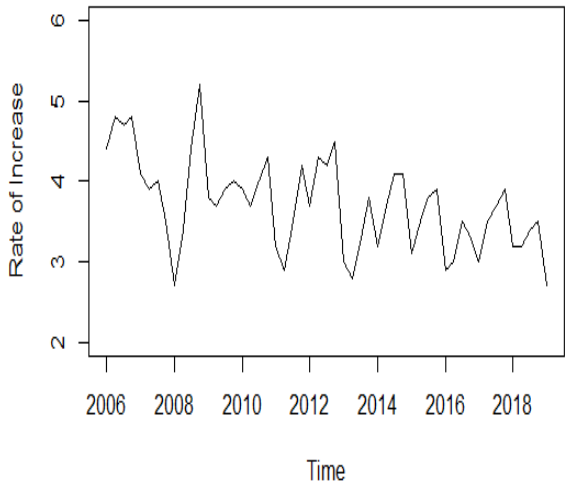
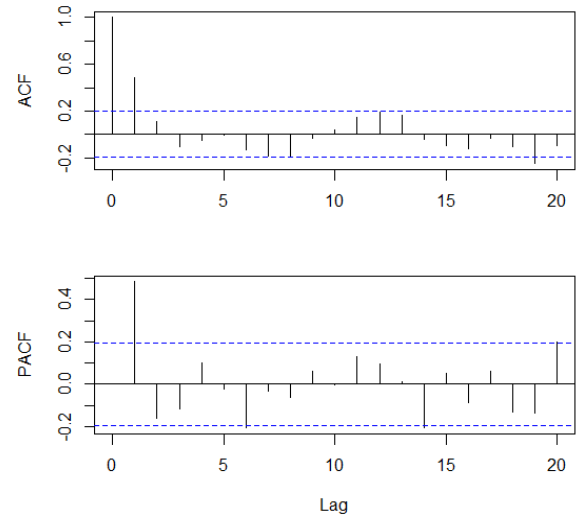

Figure 5

Figure 6

Figure 5. The time series diagram of the year-by-year growth of the primary industry.
Figure 6. The ACF and PACF of the year-by-year growth of the primary industry.

The *ARMA(1,1)* model is as follows, $\forall t \in T$:

$$Y_t = c + \phi Y_{t-1} + e_t - \theta e_{t-1},$$

where the estimated value of $e_t \sim N(0,\sigma^2)$, c is sample average $\hat{c} = 3.744$. The parameters are estimated by *algorithm 5* and we can obtain the estimated values of $\phi, \theta, \sigma$ as follows:

$$Y_t = 3.744 + 0.03925049 Y_{t-1} + e_t - 0.6014705 e_{t-1},$$

Where $e_t \sim N(0, 0.5356927^2)$.

The fitted model was evaluated and its residual satisfies the condition of independent normal distribution (see *Figure 7*). It shows that the *ARMA* model can fit the data well. Next, the model is predicted. *Table 2* gives the actual value, forecast value, relative error and confidence interval of year-by-year growth of primary industry in the quarter of 1~2,1~3,1~4 in 2018, and the prediction effect is good.

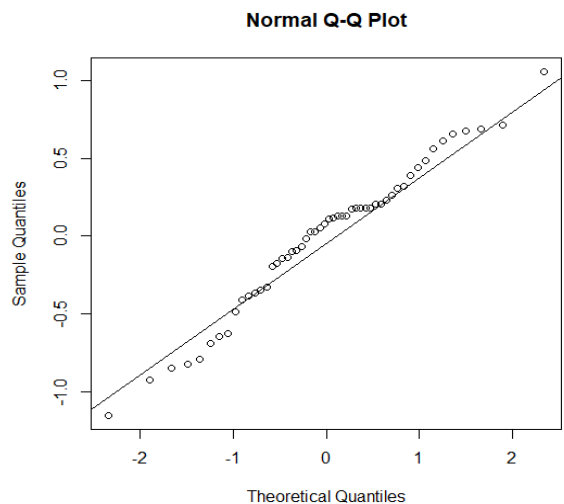

Figure 7. The Q-Q Plot fitted by ARMA model

Table 2. The prediction of the year-by-year growth of primary industry

| Time | Actual Value | Predicted Value | Relative Error | 80% Confidence Interval | 95% Confidence Interval |
|---|---|---|---|---|---|
| Q1~2,2018 | 3.2 | 3.276426 | 2.388% | (2.40,4.15) | (1.94,4.61) |
| Q1~3,2018 | 3.4 | 3.416563 | 0.487% | (2.66,4.17) | (2.26,4.57) |
| Q1~4,2018 | 3.5 | 3.438341 | 1.762% | (2.72,4.16) | (2.34,4.53) |

**Acknowledgement**

This research was financially supported by Zhejiang Provincial Natural Science Foundation of China(No. LQ18A010007).